\documentstyle[preprint,aps]{revtex}
\tightenlines
\begin{document}
\draft

\title{Remarks on vacuum fluctuations around a spinning cosmic string}


\author{
V. A. De Lorenci
\protect\thanks{Email address: \tt lorenci@cpd.efei.br}
and E. S. Moreira Jr.
\protect\thanks{Email address: \tt moreira@cpd.efei.br}}
\address{
Instituto de Ci\^encias - Escola Federal de Engenharia de Itajub\'a \\
Av. BPS 1303 Pinheirinho, 37500-903 Itajub\'a, Minas Gerais -- Brazil}
\date{September 21, 2000}
\maketitle


\begin{abstract}

We apply the point-splitting method to investigate vacuum
fluctuations in the nonglobally hyperbolic background of
a spinning cosmic string.
Implementing renormalization by removing the
Minkowski contribution it is shown that
although the Green function in the literature
satisfies the usual Green function equation,
it leads to pathological physical results.

\end{abstract}

It has been conjectured in Ref. \cite{mat90}
by using general dimensional arguments
that the zero-point energy could lead to a purely
quantum dragging effect in the background of
a spinning cosmic string.
As a first step to evaluate such a possible effect,
the ``Feynman'' Green function
$G_{{\cal F}}(x,x')$ was written in Ref. \cite{mat90}
by assuming that the standard
Schwinger proper time prescription was valid
in the context of this
nonglobally hyperbolic spacetime.

In this work the Green function in Ref. \cite{mat90} is recast
and renormalized by removing the flat spacetime contribution.
Then we use the point-splitting method
to calculate the ``vacuum fluctuations $\langle\phi^2(x)\rangle$,''
\begin{equation}
\langle\phi ^{2}(x)\rangle=
i\lim_{x'\rightarrow x}G_{{\cal F}}(x,x'),
\label{vf}
\end{equation}
for a massless
scalar field around a spinning cosmic string, showing that
it diverges at an infinite countable set of spatial points,
despite the fact that $G_{{\cal F}}(x,x')$ satisfies the
usual Green function equation.

Throughout the text
$c=\hbar=1$
and $G=1/4$.
Our notation differs slightly from that in Ref. \cite{mat90}.

The spacetime around a spinning cosmic string \cite{vil94} is
characterized by the Minkowski
line element written in cylindrical coordinates
\begin{equation}
ds^{2}=d\bar{t}^{\,2}-dr^{2}-r^{2}d\varphi^{2}-dz^{2},
\label{le}
\end{equation}
and by the identification
\begin{equation}
\left(\bar{t},r ,\varphi,  z\right)
\sim  \left(\bar{t}+2\pi S,r ,\varphi +2\pi\alpha ,  z\right),
\label{id}
\end{equation}
where the spin density and the cone parameter satisfy
$S\geq 0$ and $\alpha>0$, respectively.
By redefining the time coordinate as
$t:=\bar{t}-S\varphi /\alpha$,
Eq. (\ref{le}) becomes
$ds^{2}=(dt+Sd\varphi/\alpha)^{2}-dr^{2}
-r^{2}d\varphi^{2}-d z^{2}$
and Eq. (\ref{id}) yields
$\left(t,r ,\varphi, z\right)
\sim  \left(t,r ,\varphi +2\pi\alpha ,z\right).$
As the region for which $r<S/\alpha$ is
inhabited by closed time--like curves
the background is not globally hyperbolic
and, therefore, it is not clear whether the
usual tools of quantum field theory
make sense in this arena \cite{ful89,fro98}.

The Schwinger representation
of the ``Feynman'' Green function in Ref.\cite{mat90}
involves an integration
of the product of two Bessel functions of the first kind,
which can be evaluated \cite{gra80} leading to
the following expression for $G_{{\cal F}}(x,x)$:
\begin{eqnarray}
G_{{\cal F}}(r)= \int^\infty_{0} dt
\frac{(t/i\pi)^{1/2}}{\alpha (4\pi i t)^2}e^{-r^2/2it}
\int^{\infty}_{-\infty}d\omega\ e^{it\omega^2}
\sum_{n=-\infty}^{\infty} I_{|n+\omega S|/\alpha}(r^2/2it),
\label{2-1}
\end{eqnarray}
where $I_{\nu}$ denotes the modified Bessel function of
the first kind.
Using a convenient integral representation for $I_{\nu}$
(see expression 8.431-5 in Ref. \cite{gra80}),
it follows that
\begin{eqnarray}
\sum_{n=-\infty}^{\infty} I_{|n+\omega S|/\alpha}(z)
= \alpha e^z -\frac{1}{\pi} \int_{0}^{\infty} dx\  e^{-z\cosh x}
\sum_{n=-\infty}^{\infty} \sin (|n+\omega S|\pi/\alpha)\
e^{-|n+\omega S|x/\alpha},
\label{3-1}
\end{eqnarray}
which holds for $\alpha>1/2$, being crucial to
implementing renormalization and to
performing the
integration over $\omega$ in Eq. (\ref{2-1}).
[We should mention that expression (\ref{3-1}),
which can be extended to $\alpha\leq 1/2$,
has previously been used in related contexts \cite{shi92,pon98}.]
Thus considering Eq. (\ref{3-1}), Eq. (\ref{2-1}) becomes
\begin{eqnarray}
G_{\cal F}(r) &=& \int^\infty_{0} \frac{dt}{(4\pi i t)^2}
\nonumber \\
 &&- \frac{i^{3/2}}{16\alpha\pi^{7/2}}
\int^\infty_{0}\frac{ dt}{t^{3/2}}e^{-r^2/2 i t}
\sum_{n=-\infty}^{+\infty}
\int^{\infty}_{-\infty} d\omega\ e^{it\omega^2}
\sin (|n+\omega S|\pi/\alpha)
\nonumber\\
&&\times \int_{0}^{\infty} dx\ e^{-(r^2/2it)\cosh x}\
e^{-|n+\omega S|x/\alpha}.
\label{4-2}
\end{eqnarray}
Now the ultraviolet divergent term (the Minkowski contribution)
in Eq. (\ref{4-2}) is dropped.
By inserting
$\delta[\lambda-(n+\omega S)/\alpha]$
in Eq. (\ref{4-2}) and using Poisson's formula,
we are able to perform the integration over $\omega$ \cite{gra80}
resulting
\begin{eqnarray}
&&G_{\cal F}(r) = \frac{1}{16\pi^3}
\int^\infty_{0} \frac{dt}{t^{2}}e^{-r^2/2 i t}
\nonumber
\\
&&\times\sum_{n=-\infty}^{\infty}
\int^{\infty}_{-\infty} d\lambda\ e^{(2\pi nS)^2/4it}
e^{2\pi i \alpha n\lambda}\sin (|\lambda|\pi)
\int_{0}^{\infty} dx\ e^{-(r^2/2it)\cosh x}
e^{-|\lambda|x}.
\label{4-3}
\end{eqnarray}
The next step is to evaluate the integrations over
$\lambda$ and $t$ \cite{gra80}. Considering Eq. (\ref{vf})
it follows that
\begin{eqnarray}
&&\langle\phi^2(r)\rangle
= -\frac{1}{8\pi^2 r^2}\int^{\infty}_{0}dx
\frac{1}{(\pi^2+x^2)\cosh^2(x/2)} \nonumber\\
&&-\frac{1}{4\pi^2 r^2}\int^{\infty}_{0}dx\sum_{n=1}^{\infty}
\frac{x^2-\pi^2(4\alpha^2 n^2-1)}{\left[\pi^2(2\alpha n+ 1)^2
+ x^2\right] \left[\pi^2(2\alpha n - 1)^2 +x^2\right]
\left[\cosh^2(x/2) - (n\pi S/r)^2\right]}.
\label{phi2-sum}
\end{eqnarray}
Finally,
by considering the power series expansion
of $\psi(z)$ (the logarithmic derivative of the gamma function)
and its properties, the summation in Eq. (\ref{phi2-sum})
can be evaluated. By setting $S=0$, Eq. (\ref{phi2-sum})
reproduces the result for a spinless cosmic string in the literature
\cite{smi90},
$\langle\phi^2(r)\rangle =(\alpha^{-2} -1)/48\ \pi^{2}r^{2}$.
When $S\neq 0$, it results
\begin{eqnarray}
&&\langle\phi ^{2}(r)\rangle=
\frac{-1}{4\pi^{2}\alpha r^{2}}\int_{0}^{\infty}
\frac{dx}{B(x,y)}\times
\nonumber
\\&&
\left\{\left(\frac{1}{\pi}
\left[4\cosh^{2}(x/2)-y^{2}(\pi^{2}-x^{2})\right]
\sin(\pi/\alpha)
+2y^{2}x\sin(x/\alpha)
\right)
\frac{1}{\cosh(x/\alpha)-\cos(\pi/\alpha)}\right.
\nonumber
\\&&
\hspace{3cm}+\left.
y\left[\frac{y^{2}(\pi^{2}+x^{2})}{2\cosh(x/2)}-2\cosh(x/2)
\right]\cot\left(\frac{1}{\alpha\ y}\cosh(x/2)\right)\right\},
\label{moises}
\end{eqnarray}
where
$y:=S/\alpha r$
and $B(x,y):=4\pi^{2}y^{4}x^{2}+
[4\cosh^{2}(x/2)-y^{2}(\pi^{2}-x^{2})]^{2}$.
For an integer $n$, when $y=1/n\pi\alpha$, i.e. $r=n\pi S$,
an inspection reveals that
Eq. (\ref{moises}) diverges.

These divergences
may be indicating that it is not safe to use
standard quantum field theory prescriptions in the context of
nonglobally hyperbolic spacetimes.
Therefore
the Green function $G_{{\cal F}}(x,x')$ in Ref. \cite{mat90}
should not be used,
as in Eq. (\ref{vf}),
to investigate vacuum fluctuations.
In the light of these considerations, in order to
verify the quantum dragging effect conjecture in
Ref. \cite{mat90}, one should take the cosmic string
with a radius (thick cosmic string) bigger than
the critical radius $S/\alpha$,
such that the corresponding spacetime be globally hyperbolic
and, therefore, where standard quantum field theory
prescriptions are reliable.

Before closing, a remark is in order.
This work shows that
pathologies arise when
standard tools of second quantization
are used in the background
of a spinning cosmic string.
It is worth recalling
that standard quantum mechanics
in such a background
is also plagued by pathologies \cite{ger89}.

\acknowledgments

This work was partially supported by
{\em Conselho Nacional
de Desenvolvimento Cient\'{\i}fico e Tecnol\'ogico} (CNPq)
of Brazil.

\end{document}